\documentclass[preprint,12pt]{aastex}
\usepackage{graphicx}

\newcommand{\eg}{ e.g.}
\newcommand{\msun}{\ensuremath{\mathrm{M}_{\odot}}}
\newcommand{\nuc}[2]{\ensuremath{\mathrm {^{#2}#1}}}
\newcommand{\ttt}[1]{\ensuremath{\times 10^{#1}}}

\begin{document}

\title{Nuclear Reaction Rate Uncertainties and their effects on Nova 
Nucleosynthesis Modeling}

\author{W.~Raphael~Hix\altaffilmark{1,2}, 
Michael~S.~Smith\altaffilmark{2}, 
Anthony~Mezzacappa\altaffilmark{2,1}, 
Sumner~Starrfield\altaffilmark{3}, 
Donald~L.~Smith\altaffilmark{4}}

\altaffiltext{1}{Department of Physics \& Astronomy, University of 
Tennessee, Knoxville, TN 37996-1200}
\altaffiltext{2}{Physics Division, Oak Ridge National Laboratory, Oak
Ridge, TN 37831-6354 \\
Managed by UT-Battelle, LLC, for the U.S. Department of Energy  
under contract DE-AC05-00OR22725}
\altaffiltext{3}{Department of Physics \& Astronomy, Arizona State University, 
Tempe, AZ 85287-1504}
\altaffiltext{4}{Technology Development Division, Argonne National Laboratory,
Argonne, IL 60439}

\begin{abstract}
The nucleosynthesis and other observable consequences of a nova
outburst depend sensitively on the details of the thermonuclear
runaway which initiates the outburst.  One important source of
uncertainty in our current models is the nuclear reaction data used as
input for the evolutionary calculations.  We present preliminary
results of the first analyses of the impact on nova nucleosynthesis
of all reaction rate uncertainties considered simultaneously.
\end{abstract}

\section{Nuclear Uncertainties in Novae}

Observations of nova outbursts have revealed an elemental composition that
differs markedly from solar.  Theoretical studies indicate that these
differences are caused by the combination of convection with explosive
hydrogen burning which results in a unique nucleosynthesis that is rich in
odd-numbered nuclei such as $\nuc{C}{13}$, $\nuc{N}{15}$, and $\nuc{O}{17}$. 
Such nuclei are difficult to form in other astrophysical hosts.  Many of
the proton-rich nuclei produced in nova outbursts are radioactive, offering
the possibility of direct observation with $\gamma$-ray instruments. 
Potentially important $\gamma$-ray sources include \nuc{Al}{26},
\nuc{Na}{22}, \nuc{Be}{7}, and \nuc{F}{18}.

The observable consequences of a nova outburst depend sensitively on
the details of the thermonuclear runaway which initiates the outburst. 
One important source of uncertainty is the nuclear reaction data used
as input for the evolutionary calculations \citep{STWS98}.  A number
of features conspire to magnify the effects of nuclear uncertainties
on nova nucleosynthesis.  First, the similarity of the nuclear burning
and convective timescales in novae results in nuclear burning which is
far from the steady state which typifies quiescent burning.  Further,
many reactions of relevance to novae involve unstable proton-rich
nuclei, making experimental rate determinations difficult.  Finally,
for hydrodynamic conditions typical of novae, many rates depend
critically on the properties of a few individual resonances, resulting
in wide variation between different rate determinations.  As a result,
statistical model (Hauser-Feshbach) calculations, which are employed
with great success in other venues for a large number of reactions
\citep{RaTK97a}, are unreliable for many rates under nova conditions. 
As we will demonstrate, these factors greatly restrict our ability to
make precise predictions of the nucleosynthesis of novae.

\section{Monte Carlo Estimates of Uncertainties}

Though analysis of the impact of variations in the rates of a few
individual reactions has recently been performed using one-dimensional
hydrodynamic models \citep{JoCH99}, analysis of the impact of the
complete set of possible reaction rate variations in such hydrodynamic
models remains computationally prohibitive.  We therefore begin by
examining in detail the nucleosynthesis of individual zones, using
hydrodynamic trajectories (temperature and density as a function of
time) drawn from nova outburst models.  Such post-processing
nucleosynthesis simulations have been a common means of estimating
nova nucleosynthesis \citep[see \eg,][]{HiTh82,WGTR86}.  For this
presentation we are using a hydrodynamic trajectory for an inner zone
of a $1.25 \msun$ ONeMg WD, similar to that described by
\cite{PSTW95}.  These calculations were performed using a nuclear
network with 87 species, composed of elements from n and H to S,
including all isotopes between the proton drip line and the most
massive stable isotope.

To investigate the extent to which nuclear reaction uncertainties
translate into abundance differences, we use a Monte Carlo technique
which simultaneously assigns a random enhancement factor to each
reaction rate in the nuclear network.  Monte Carlo methods have been
employed with success in the analysis of Big Bang nucleosynthesis
\citep{SmKM93}, but this is their first application to other
thermonuclear burning environments.  This is also the first analysis
of nova nucleosythesis to examine the impact of all the reaction
uncertainties simultaneously.  The reaction rate enhancement factors
are distributed according to the log-normal distribution, which is the
correct uncertainty distribution for quantities like reaction rates
which are manifestly positive \citep{Smit91}.  The probability density
function is given by
\begin{equation}
    p_{log-normal}(x) = \frac{1}{\sqrt{2 \pi} \beta x}
    \exp \left(- \frac{(\ln x - \alpha)^2}{2 \beta^2} \right) \ ,
    \label{eq:lognorm}
\end{equation}
where $\alpha$ and $\beta$ are the (logarithmic) mean and standard
deviation.  Unlike the normal (Gaussian) distribution, where
confidence intervals are symmetric in the additive sense, the
confidence intervals for the log-normal distribution are symmetric in
a multiplicative sense.  For small uncertainties $(<20\%)$, the
difference between the log-normal distribution and the normal
distribution is small.  However, for uncertainties of larger sizes,
such as those encountered in this problem, the difference is
important.  For example, assuming a normal distribution with a
relative uncertainty of 50\% implies that there is a 5\% chance of a
negative reaction rate.  

The choice of uncertainty for each reaction rate is crucial for this
analysis.  For $\beta$-decays, we have used the experimental
uncertainties \citep{NuWC00}.  For many of the other rates important
for nova nucleosynthesis, the uncertainties in the reaction rate are
unknown or are likely to be dominated by systematic effects such as
missing resonances.  We have therefore assigned uncertainties by
category.  Rates calculated by Hauser-Feshbach methods as well as
rates whose measurement require radioactive ion beams were assigned
uncertainties of $\sim 50\% \ (\beta=\ln(1.5))$.  For rates measured
with stable beams, we have assigned $\beta=\ln(1.2)$.  Note that
relatively few reactions, especially among unstable nuclei, have
measurement uncertainties this small.  We have chosen these small 
uncertainties to obtain a conservative estimate of the impact of 
nuclear uncertainties on nucleosynthesis predictions.

\begin{figure}[tb]
  \centering
  \includegraphics[width=\textwidth]{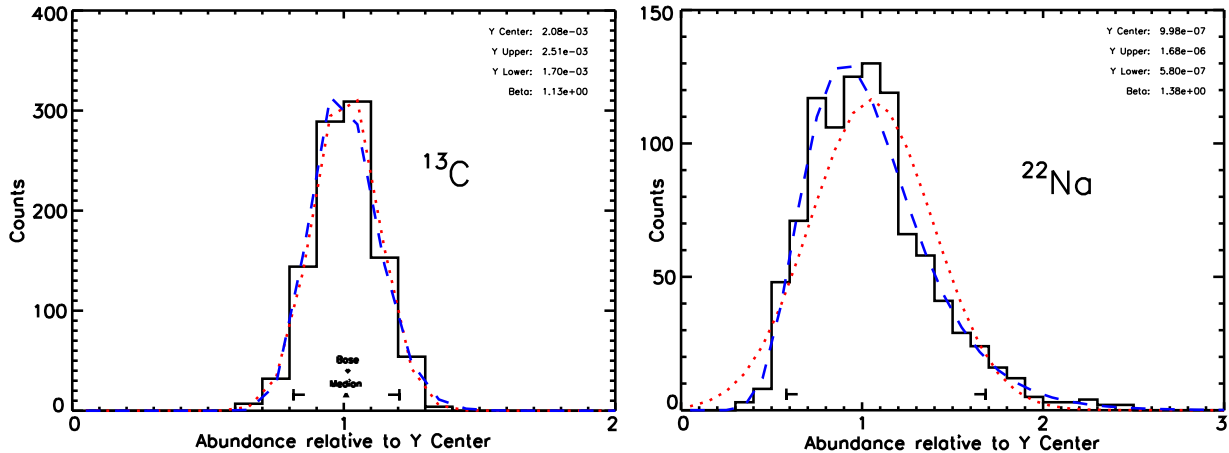}
  \caption{Histograms showing relative deviations in abundance.  Y
    Center, Y Upper, and Y Lower correspond, respectively, to the
    central abundance and the upper and lower error bars from
    Fig.~\ref{fig:mcabund}.  The dotted curves are normal distributions
    with the same (arithmetic) mean and standard deviation as the Monte
    Carlo distribution.  The dashed curves are log-normal distributions
    with (logarithmic) mean and standard deviation from the Monte
    Carlo.}
  \label{fig:hist}
\end{figure}

The histograms drawn in Figure~\ref{fig:hist} show the abundance
distributions which result from the Monte Carlo analysis (992 trials)
for two representative nuclei, \nuc{C}{13} and \nuc{Na}{22}.  From
distributions such as these, mean values and 90\% confidence limits
are extracted.  The normal and log-normal distributions with widths
drawn from these Monte Carlo abundance distributions are also plotted
in Fig.~\ref{fig:hist}.  Note the similarity between the normal and
log-normal distributions for \nuc{C}{13}, which has a standard
deviation of 12\%, while these distributions differ markedly for
\nuc{Na}{22}, which has a standard deviation of 32\%.

\begin{figure}[ttb]
  \centering
  \includegraphics{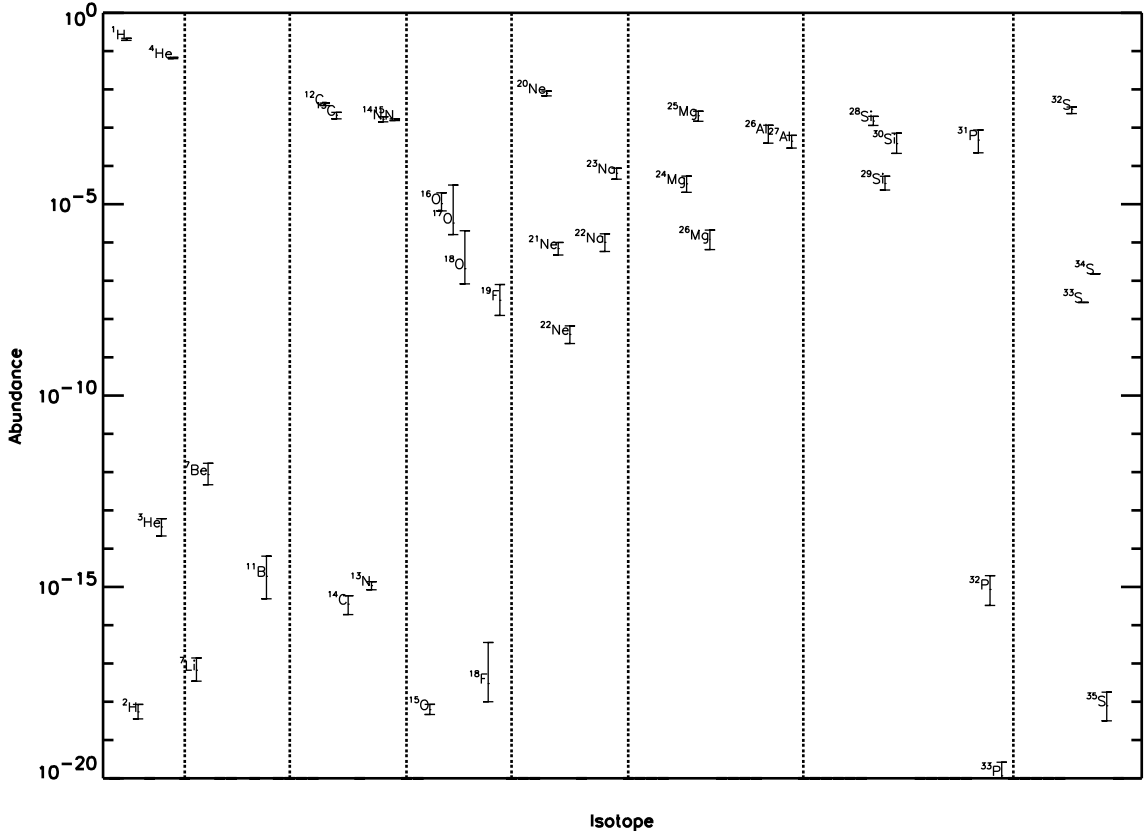}
  \caption{Final Abundances (elapsed time = $4.7\ttt{5} \sec$ after peak).}
  \label{fig:mcabund}
\end{figure}

Figure~\ref{fig:mcabund} shows the mass fractions of each species at
the end of the simulation, $4.7\ttt{5} \sec$ after peak temperature. 
Because of the long time which has elapsed, the unstable proton-rich
nuclei have decayed, reducing their abundances to less than
$10^{-20}$.  The error bars displayed in Fig.~\ref{fig:mcabund} are
the 90\% confidence intervals which result from the Monte Carlo
analysis.  As evidenced by the error bars in Fig.~\ref{fig:mcabund},
the impact of even our conservatively chosen variations in reaction
rates on the nucleosynthesis is large.  While broader conclusions will
require similar analysis of an entire model, a number of interesting
points can be drawn from the analysis of this single hydrodynamic
trajectory.  First, for reaction rate variations of the size 
considered here, the impact on the rate of energy production is
small.  At the 90\% confidence level, variations in the amount of
hydrogen consumed and helium produced represent $\sim 10\%$ variation
in the thermonuclear energy released.  This is essential if
post-processing techniques are to prove useful.  Second, for the most
abundant CNO isotopes, such as \nuc{C}{13} and \nuc{N}{15}, 2$\sigma$
variations of 5-20\% are common, though some less abundant CNO
isotopes show larger variations.  For example, the 90\% confidence
interval for the mass fraction of \nuc{O}{17} has a lower limit of
$2.7 \ttt{-5}$ and an upper limit of $5.3\ttt{-4}$, a span of a factor
of 20.  Third, for many of the nuclei in the NeNa and MgAl cycles,
including the $\gamma$-ray source nuclei \nuc{Na}{22} and
\nuc{Al}{26}, the 90\% confidence interval spans variations of a
factor of 3.  This represents nearly a factor of 2 difference in the
distance from which novae may be observable with $\gamma$-ray
telescopes.

\section{Determining the impact of individual reactions on the 
nucleosynthesis of individual isotopes}

\begin{figure}[p]
    \centering
    \includegraphics{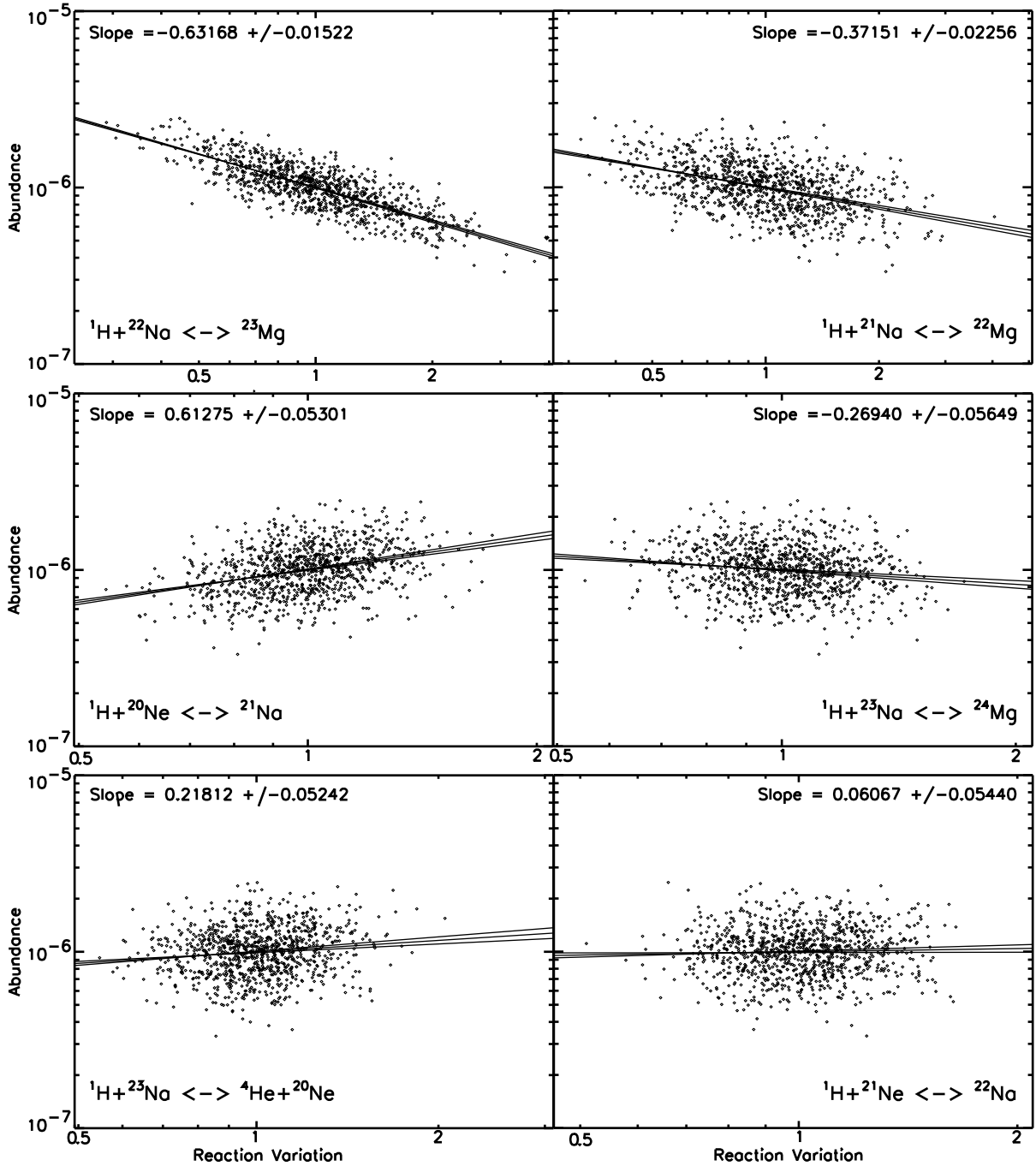}
%   \includegraphics[angle=90,width=.48\textwidth]{cor_O125_M_Na22_123.ps}
%   \includegraphics[angle=90,width=.48\textwidth]{cor_O125_M_Na22_120.ps} 
%   \par
%   \includegraphics[angle=90,width=.48\textwidth]{cor_O125_M_Na22_110.ps}
%   \includegraphics[angle=90,width=.48\textwidth]{cor_O125_M_Na22_126.ps}
%   \par
%   \includegraphics[angle=90,width=.48\textwidth]{cor_O125_M_Na22_300.ps}
%   \includegraphics[angle=90,width=.48\textwidth]{cor_O125_M_Na22_112.ps}
    \caption{Correlations of variations of individual reactions with the 
    abundances of \nuc{Na}{22}  The three lines portray the most 
    probable values of the correlation bounded by its 90\% confidence 
    limits.}
    \label{fig:corr}
\end{figure}

While the use of the Monte Carlo described in the previous section
allows us to estimate the uncertainty in our nucleosynthetic
calculations caused by nuclear reaction uncertainties, it does not
indicate which reactions influence the production of an individual
isotope.  In order to determine this connection, we re-examine our
Monte Carlo trials to find correlations between the abundance variations
and the variations in all of the included reactions.  Given the wide
range of variations, we test for correlations in the logarithm of
the abundance change with respect to the logarithm of the reaction
variations.  Figure~\ref{fig:corr} displays some representative
correlations between reactions rates and the abundance of
\nuc{Na}{22}.  Since we simultaneously vary all the included reactions
rates, the resultant plots show some scatter, however statistically
significant correlations are still achieved for a number of reactions. 
Fig.~\ref{fig:corr}a, Fig.~\ref{fig:corr}b, and Fig.~\ref{fig:corr}d
show that increases in $\nuc{Na}{21}(p,\gamma)\nuc{Mg}{22}$,
$\nuc{Na}{22}(p,\gamma)\nuc{Mg}{23}$, and
$\nuc{Na}{23}(p,\gamma)\nuc{Mg}{24}$ resulted in reduced abundances of
\nuc{Na}{22}, while Fig.~\ref{fig:corr}c and Fig.~\ref{fig:corr}e show
that increases in the reaction rate of
$\nuc{Ne}{20}(p,\gamma)\nuc{Na}{21}$ and
$\nuc{Na}{23}(p,\alpha)\nuc{Ne}{20}$ increased the abundance of
\nuc{Na}{22}.  Interestingly, Fig.~\ref{fig:corr}f demonstrates that, 
for reaction rate variations of the size considered here, the
abundance of \nuc{Na}{22} is essentially independent of the rate of
$\nuc{Ne}{21}(p,\gamma)\nuc{Na}{22}$, even though this reaction in the
primary production channel for \nuc{Na}{22}.  Much larger variations 
in the reaction rate of $\nuc{Ne}{21}(p,\gamma)\nuc{Na}{22}$ would 
necessitate reexamination of this conclusion. 

\cite{JoCH99} have recently performed a set of spherically symmetric
nova outburst simulations in which they vary individual reaction rates
in order to examine the impact of these rates on the production of
\nuc{Na}{22} and \nuc{Al}{26}.  They found that variations in the
rates of $\nuc{Na}{21}(p,\gamma)\nuc{Mg}{22}$ and
$\nuc{Na}{22}(p,\gamma)\nuc{Mg}{23}$ produced significant variations
in the \nuc{Na}{22} abundance, with
$\nuc{Na}{23}(p,\gamma)\nuc{Mg}{24}$ having a smaller effect.  Our
results confirm these conclusions, however we find that the reaction
$\nuc{Ne}{20}(p,\gamma)\nuc{Na}{21}$ has effects similar in strength
to the former two and that $\nuc{Na}{23}(p,\alpha)\nuc{Ne}{20}$ has an
effect similar in strength to the latter.  Examination of these rates
within hydrodynamic simulations of novae would present an excellent
test of the predictive power of the Monte Carlo analysis.

\section{Conclusions}

We have performed analysis of the effects of reaction rate
uncertainties on nova nucleosynthesis.  Our Monte Carlo analysis is
the first such analysis that simultaneously examines the impact of all
of the uncertainties.  The large uncertainties in nova nucleosynthesis
that we find constrain our ability to make detailed comparisons
between theoretical models for the nova outburst and astrophysical
observations to a degree which is often ignored.  Further, knowledge
gleaned from this analysis of which reaction rate uncertainty
dominates the uncertainty in a given isotope helps prioritize new
experimental measurements.  Only with improved knowledge of these
uncertain rates, both experimental and theoretical, can we provide
tight constraints on the nova outburst from its nucleosynthetic
products.

%\bibliographystyle{apj}
%\bibliography{apjmnemonic,astro}

\begin{thebibliography}{9}
\expandafter\ifx\csname natexlab\endcsname\relax\def\natexlab#1{#1}\fi

\bibitem[{{Hillebrandt} \& {Thielemann}(1982)}]{HiTh82}
{Hillebrandt}, W. \& {Thielemann}, F.-K. 1982, ApJ, 255, 617

\bibitem[{{Jos\'e} {et~al.}(1999){Jos\'e}, {Coc}, \& {Hernanz}}]{JoCH99}
{Jos\'e}, J., {Coc}, A., \& {Hernanz}, M. 1999, ApJ, 520, 347

\bibitem[{{Politano} {et~al.}(1995){Politano}, {Starrfield}, {Truran}, {Weiss},
  \& {Sparks}}]{PSTW95}
{Politano}, M., {Starrfield}, S., {Truran}, J.~W., {Weiss}, A., \& {Sparks},
  W.~M. 1995, ApJ, 448, 807

\bibitem[{{Rauscher} {et~al.}(1997){Rauscher}, {Thielemann}, \&
  {Kratz}}]{RaTK97a}
{Rauscher}, T., {Thielemann}, F.-K., \& {Kratz}, K.-L. 1997, Nucl. Phys. A,
  621, 331

\bibitem[{Smith(1991)}]{Smit91}
Smith, D.~L. 1991, Probability, Statistics, and Data Uncertainties in Nuclear
  Science and Technology (LaGrange Park: Am. Nuc. Soc.)

\bibitem[{{Smith} {et~al.}(1993){Smith}, {Kawano}, \& {Malaney}}]{SmKM93}
{Smith}, M.~S., {Kawano}, L.~H., \& {Malaney}, R.~A. 1993, ApJS, 85, 219

\bibitem[{{Starrfield} {et~al.}(1998){Starrfield}, {Truran}, {Wiescher}, \&
  {Sparks}}]{STWS98}
{Starrfield}, S., {Truran}, J.~W., {Wiescher}, M.~C., \& {Sparks}, W.~M. 1998,
  MNRAS, 296, 502

\bibitem[{Tuli(2000)}]{NuWC00}
Tuli, J.~K. 2000, Nuclear Wallet Cards

\bibitem[{{Wiescher} {et~al.}(1986){Wiescher}, {Gorres}, {Thielemann}, \&
  {Ritter}}]{WGTR86}
{Wiescher}, M., {Gorres}, J., {Thielemann}, F.-K., \& {Ritter}, H. 1986, A\&A,
  160, 56

\end{thebibliography}

\end{document}